\documentclass{elsart41}

\usepackage{graphicx}          
\usepackage{amsmath}           
\usepackage{amssymb}           

\begin{document}

\begin{frontmatter}
\title{Coherent transport of interacting electrons through a single scatterer}

\author{Martin Mo\v{s}ko},
\author{Pavel Vagner},
\author{Andrej Gendiar},
\author{Radoslav N\'{e}meth}

\address{Institute of Electrical Engineering, Slovak Academy of
Sciences, 84104~Bratislava, Slovakia}

\begin{abstract}
Using the self-consistent Hartree-Fock method, we calculate the
persistent current of weakly-interacting spinless electrons in a
one-dimensional ring containing a single $\delta$ barrier. We find
that the persistent current decays with the system length ($L$)
asymptotically like $I \propto L^{-1-\alpha}$, where $\alpha > 0$
is the power depending only on the electron-electron interaction.
We also simulate tunneling of the weakly-interacting
one-dimensional electron gas through a single $\delta$ barrier in
a finite wire biased by contacts. We find that the Landauer
conductance decays with the system length asymptotically like
$L^{-2\alpha}$. The power laws $L^{-1-\alpha}$ and $L^{-2\alpha}$
have so far been observed only in correlated models. Their
existence in the Hartree-Fock model is thus surprising.
\end{abstract}

\end{frontmatter}

%


Magnetic flux applied through the opening of a mesoscopic
conducting ring gives rise to a persistent electron current
circulating along the ring \cite{Imry-book}. Here we study the
persistent current of interacting spinless electrons in a
one-dimensional (1D) ring with a single scatterer.

For non-interacting electrons the persistent current ($I$) depends
on the magnetic flux ($\phi$) and ring length ($L$) as
\cite{Gogolin}
\begin{equation} \label{I-nonint-approx}
I = (ev_F/2 L) |\tilde{t}_{k_{F}}| \sin(\phi'),
\end{equation}
if $|\tilde{t}_{k_{F}}| \ll 1$. In eq. \eqref{I-nonint-approx}
$\phi' \equiv 2\pi \phi/\phi_0$, $\phi_0 = h/e$ is the flux
quantum, $\tilde{t}_{k}$ is the electron transmission amplitude
through the scatterer, $k_{F}$ is the Fermi wave vector, and $v_F$
is the Fermi velocity. For a repulsive electron-electron
interaction the spinless persistent current was derived in the
Luttinger liquid model \cite{Gogolin}. For $L \rightarrow \infty$
\begin{equation} \label{I-Luttinger}
I \propto L^{-\alpha-1} \sin(\phi'),
\end{equation}
where $\alpha>0$ depends only on the e-e interaction.

In this work we find similar results in the Hartree-Fock model. We
consider $N$ interacting 1D electrons with free motion along a
circular ring threaded by magnetic flux $\phi=BS=AL$, where $S$ is
the area of the ring, $B$ is the magnetic field threading the
ring, and $A$ is the magnitude of the vector potential. In the
Hartree-Fock model the many-body wave function is the Slater
determinant of single-electron wave functions $\psi_k(x)$. These
wave functions obey the Hartree-Fock equation
\begin{multline}\label{Eqs-Schroding-2}
\bigg[
\frac{\hbar^2}{2m}\left(-i\frac{\partial}{\partial x}+
\frac{2\pi}{L}\frac{\phi}{\phi_0}\right)^2 +
\gamma\delta(x)
\\
+U_H(x)+U_F(k,x) \bigg] \psi_k(x) = \varepsilon_k\psi_k(x)
\end{multline}
with cyclic boundary condition $\psi_k(x+L)=\psi_k(x)$, where $m$
is the electron effective mass, $x$ is the electron coordinate
along the ring, $\gamma\delta(x)$ is the potential of the
scatterer, the Hartree potential is given by
\begin{equation}\label{Eqs-Hartree}
  U_H(x)=\sum_{k'} \int dx' V(x-x')|\psi_{k'}(x')|^2,
\end{equation}
the Fock term is written as an effective potential
\begin{multline}\label{Eqs-Fock}
  U_F(k,x)= -\frac{1}{\psi_k(x)}
\\ \times
\sum \limits_{k'}
        \int dx' V(x-x')\psi_k(x')\psi_{k'}^*(x')\psi_{k'}(x),
\end{multline}
and $V(x-x')$ is the electron-electron (e-e) interaction.

We solve equation \eqref{Eqs-Schroding-2} coupled with the
potentials \eqref{Eqs-Hartree} and \eqref{Eqs-Fock} using
self-consistent numerical iterations \cite{Nemeth}. We obtain
numerically the single-particle states $\psi_k(x)$ and
$\varepsilon_k$, the energy of the Hartree-Fock groundstate, $E$,
and eventually the persistent current $I= - \partial
E(\phi)/\partial \phi$.

We present results for the GaAs ring with electron density $n=5
\times 10^7$ m$^{-1}$, effective mass $m=0.067$~$m_0$, and e-e
interaction
\begin{equation} \label{VeeExp}
V(x - x') = V_0 \,  e^{- \left| x - x' \right|/d},
\end{equation}
where $V_0 = 34$~meV and $d = 3$ nm. The interaction
\eqref{VeeExp} is short-ranged. It emulates screening and allows
comparison with correlated models \cite{Gogolin,Meden} which also
use the e-e interaction of finite range.

We study rings with a strong scatterer ($|\tilde{t}_{k_{F}}| \ll
1$), for which the asymptotic behavior with $L$ is reachable for
not too large $L$ \cite{Meden}. To show results typical of
$|\tilde{t}_{k_{F}}| \ll 1$, we use the $\delta$ barrier with
transmission $|\tilde{t}_{k_{F}}| = 0.03$.

\begin{figure}[t]
\centerline{\includegraphics[clip,width=0.95\columnwidth]{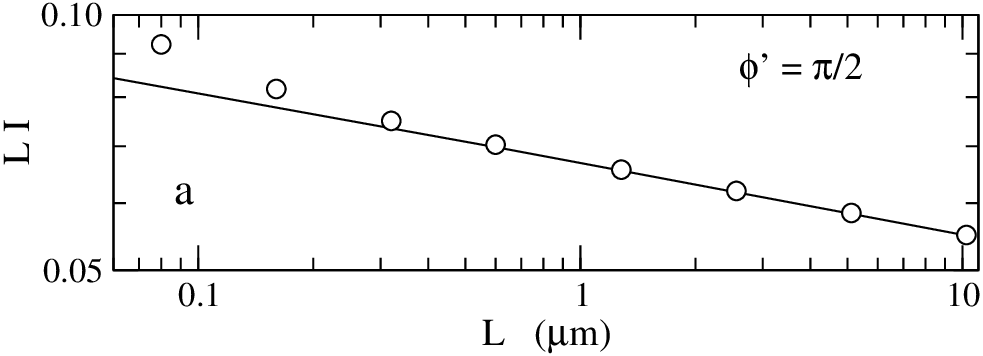}}
\centerline{\includegraphics[clip,width=\columnwidth]{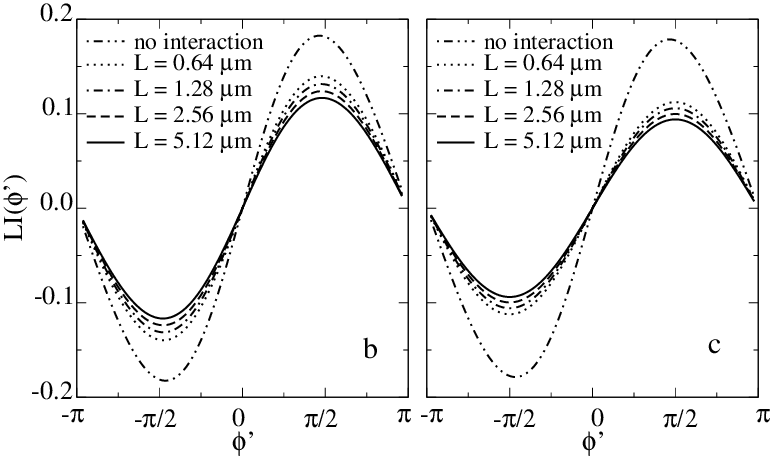}}
\caption{Panel \emph{a}: Persistent current $LI(\phi' = \pi/2)$
versus $L$. The Hartree-Fock data (open circles) follow for large
$L$ the scaling law $LI \propto L^{-\alpha}$ shown in a full line,
with $\alpha=0.0855$ obtained as discussed in the text. Panels
\emph{b} and \emph{c}: Persistent current $LI(\phi')$ for various
$L$; panel \emph{b} shows the Hartree-Fock results while panel
\emph{c} shows the scaling law of the Luttinger liquid. All data
are normalized by the constant $ev_{F}/2$.} \label{Fig:power law}
\end{figure}

Panel \emph{a} of figure \ref{Fig:power law} shows in log scale
the persistent current $LI(\phi' = \pi/2)$ as a function of $L$.
The full line is the power law $LI \propto L^{-\alpha}$ predicted
by equation \eqref{I-Luttinger}. For weak e-e interaction ($\alpha
\ll 1$) it holds \cite{Matveev-93} that $\alpha =
[V(0)-V(2k_F)]/2\pi \hbar v_F$, where $V(q)$ is the Fourier
transform of the e-e interaction $V(x-x')$. The Fourier transform
of our interaction \eqref{VeeExp} reads $V(q) = 2V_0
d/(1+q^2d^2)$. For the above parameters $\alpha = 0.0855$. The
full line fits our Hartree-Fock data for large $L$.

Panel  \emph{b} shows our Hartree-Fock results for $LI(\phi')$. To
compare our results with the scaling law \eqref{I-Luttinger}, we
formulate the relation \eqref{I-Luttinger} as follows
\cite{Gogolin}. We replace the bare transmission amplitude
$\tilde{t}_{k_F}$ in the non-interacting scaling law
(\ref{I-nonint-approx}) by the transmission amplitude of the
correlated electron gas, $t_{k_F} \simeq \tilde{t}_{k_{F}}
\left(d/L\right)^\alpha$, which holds \cite{Matveev-93} for small
$\tilde{t}_{k_F}$. We obtain the scaling law \eqref{I-Luttinger}
including the proportionality constant $const = e v_F \left|
\tilde{t}_{k_F} \right| d^{\alpha}/2$. This scaling law is
presented in panel \emph{c}. It can be seen that the results of
panels \emph{b} and \emph{c} are in good accord.

Finally, the Hartree-Fock equation \eqref{Eqs-Schroding-2} can be
used to study the conductance of a straight 1D wire biased by
contacts, if we omit the term $\propto \phi$. Of course, we also
replace the cyclic boundary condition by the boundary conditions
of the tunneling problem \cite{Matveev-93},
\begin{equation} \label{BCondgt}
\psi_k(x=-L/2)=e^{ikx}+r_k e^{-ikx}, \psi_k(x=L/2)=t_k e^{ikx},
\end{equation}
where $r_k$ is the reflection amplitude and $t_k$ is the
transmission amplitude (analogously for the electrons entering the
wire from the right). We have solved this Hartree-Fock problem
self-consistently and we have evaluated the Landauer conductance
$(e^2/h) |t_{k_F}|^2$.

\begin{figure}[t]
\centerline{\includegraphics[clip,width=\columnwidth]{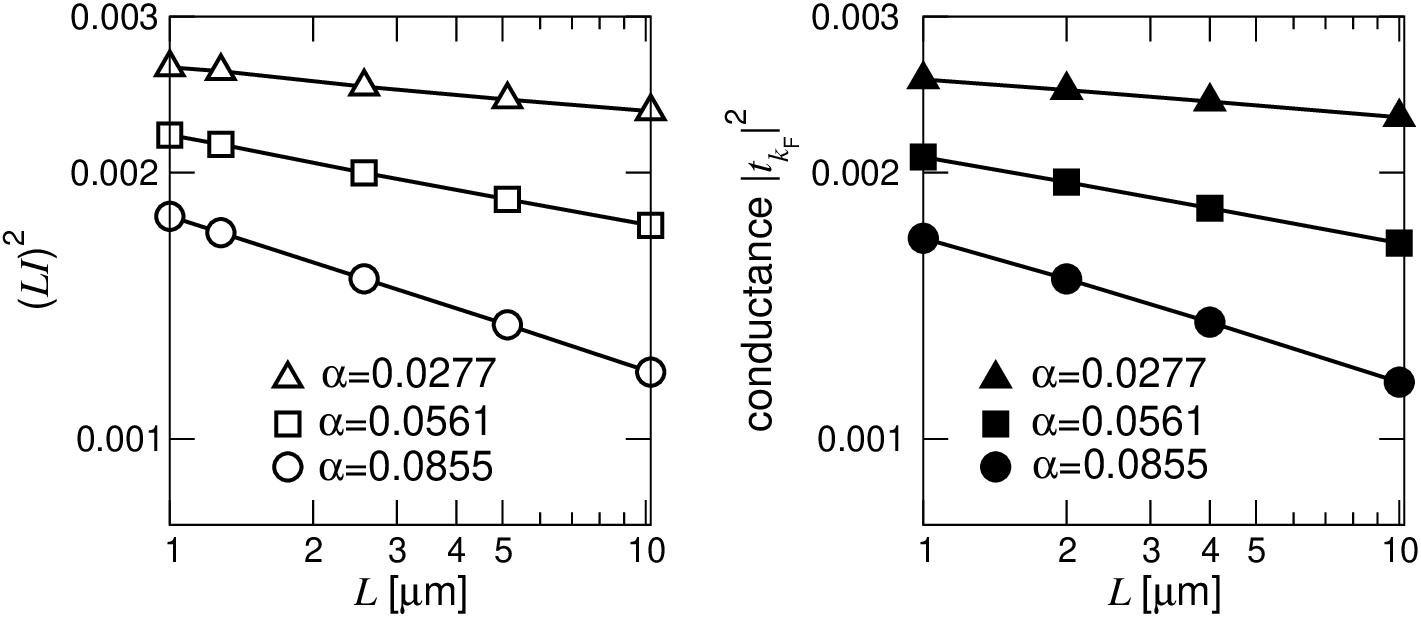}}
\caption{Left: Square of the persistent current $LI$ as a function
of $L$ for $\phi' = \pi/2$ and for various e-e interaction
strengths $\alpha$. Right: Landauer conductance $|t_{k_F}|^2$ of
the equivalent 1D wire.
 } \label{Fig:power-law2}
\end{figure}

The result is shown in figure \ref{Fig:power-law2} together with
the square of $LI$ for the equivalent ring. The conductance scales
like $L^{-2\alpha}$ and so does the square of $LI$.

\begin{figure}[t]
\centerline{\includegraphics[clip,width=\columnwidth]{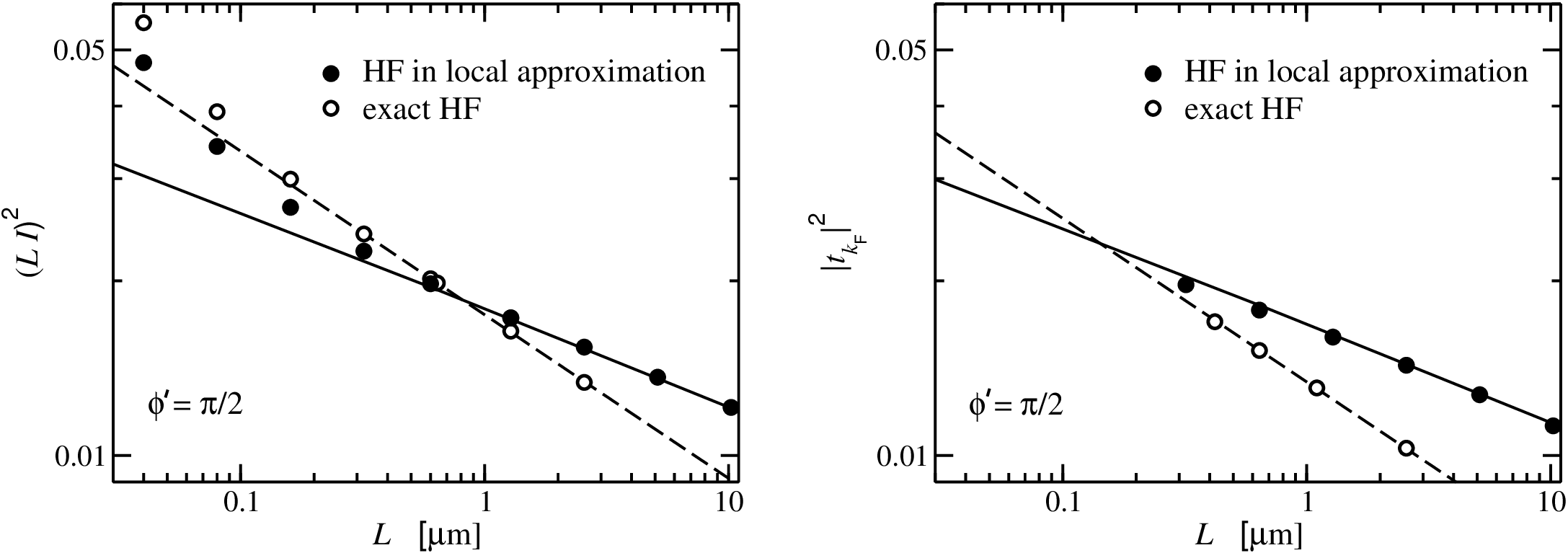}}
\caption{Left: Square of the persistent current $LI$ as a function
of $L$ for $\phi' = \pi/2$ and $\alpha=0.0855$. Right: Landauer
conductance $|t_{k_F}|^2$ of the equivalent 1D wire.
 } \label{Fig:power-law3}
\end{figure}

We thank for the APVT grant APVT-51-021602.

%
%

\end{document}